\documentclass[12pt]{article}
\usepackage[cp1251]{inputenc}
\usepackage{epsf}

\title{Reanalysis of QCD sum rules for nucleon with perturbative correction.}
\author{A.G.Oganesian\\
Institute of Theoretical and
Experimental Physics,\\
B.Cheremushkinskaya 25, 117218 Moscow,Russia}
\date{}
\begin{document}
\maketitle

\newcommand{\be}{\begin{equation}}
\newcommand{\ee}{\end{equation}}

\def\la{\mathrel{\mathpalette\fun <}}
\def\ga{\mathrel{\mathpalette\fun >}}
\def\fun#1#2{\lower3.6pt\vbox{\baselineskip0pt\lineskip.9pt
\ialign{$\mathsurround=0pt#1\hfil##\hfil$\crcr#2\crcr\sim\crcr}}}

\begin{abstract}
A new analysis of baryon sum rules is done with modern values of
parameters and known perturbative corrections. The restriction for
gluon and quark condensates  and the new value of nucleon coupling
constant are found.
\end{abstract}

\vspace{5mm}

\vspace{1cm}

\begin{center}
{\bf 1.Introduction}
\end{center}

\vspace{5mm}

During last 20 years in the framework of QCD sum rules the large
number of different hadron properties (both static and dynamic, as
masses, various decay widths, structure functions and so on) was
calculated in non-model way. It is well known that the only
parameter in this approach are values of vacuum average of the
operators appeared in OPE, like gluon condensate $ b=(2\pi)^2
\langle (\alpha/\pi) G^n_{\mu \nu}G^n_{\mu \nu} \rangle$, quark
condensate $a=-(2\pi)^2(\langle \bar{\psi}\psi \rangle$ e.t.c.
Except this parameters, from consideration of few simple
correlator with two hadronic currents, (which usually are used to
determine hadron masses), also the so-called coupling constants of
hadronic current $g^h$ defined as $g^h \sim\langle 0 \vert j^h
\vert h \rangle$ are fixed. This coupling constants appear in a
large number of other sum rules, where analogous  hadron current
is used. For example in the case of nucleon such "basic" role
plays sum rule for on 2-point correlator

\be \Pi =\int e^{iqx} d^4 x \langle 0 \mid T \eta (x)
\overline{\eta} (0) \mid 0 \rangle \ee

where $\eta$ is a current with proton quantum number. In what
following we will choose it as

\be \eta = \varepsilon^{abc} (u^aC\gamma^{\lambda} u^b)\gamma^5
\gamma^{\lambda}d^c \ee

This correlator (with various choices of the nucleon current) was
first investigated in 1982 in the paper $\cite{1}$, where coupling
constant $\lambda_N$, defined as $\langle 0 \mid \eta \mid p
\rangle = \lambda_N v (q) $ was found. (Here $v(q)$ is the proton
spinor $(\hat{q} - m)v (q) =0 $). Let us discuss this case more
detail, because just the reanalysis of this sum rules with account
of new theoretical and experimental results obtained since 1982 is
our purpose.

\vspace{1cm}

\begin{center}
{\bf 2. Baryon sum rule}
\end{center}

\vspace{3mm}

First the sum rules for barion from correlator (1) was found by
$\cite {1}$, a little later mistakes was corrected in $\cite {2}$.
One should note that authors of $\cite {1}$ used different choices
of proton current, but later they come to conclusion that the
choose we noted above (eq.(2)) is optimal (see $\cite {2}$). Now
it is widely used, so we also will discuss sum rules for this
choice. The procedure of deriving sum rules is standard (and at
nowadays well-known). From one side correlator (1) was calculated
and OPE terms up to dimension $d=9$ are taken into account, from
another it was saturated by physical resonances plus continuum,
then equating this two representations the following sum rules
were obtained $\cite {2}$:

$$ M^6 E_2 (s_0/M^2)L^{-4/9} + \frac{4}{3} a^2 L^{4/9}
+\frac{1}{4} bM^2 E_0(s_0/M^2) L^{-4/9} - $$

\be
 -\frac{1}{3} a^2 m^2_0/M^2 = \overline{\lambda}^2_N
exp(-m^2/M^2)\ee

\be
2a M^4 E_1(s_0/M^2) + \frac{272}{81} \frac{\alpha_s a^3}{\pi M^2}
- \frac{1}{12} ab = m\overline{\lambda}^2_N exp (-m^2/M^2)\ee

where $$E_2(x) = 1-(1+x+x^2/2)e^{-x},~~~E_1=1-(1+x)e^{-x}, ~~~E_0
= 1-e^{-x},$$

$ L =\frac{\alpha_s(\mu^2)}{\alpha_s(M^2)}$, $\overline{\lambda}^2
= 32\pi^4 \lambda^2_N$, $m$ is proton mass and quark condensate
$a$ and gluon condensate $b$ was defined above. Two sum rules
(3),(4) correspond to the amplitudes at kinematical structures
$\hat{p}$ and $I$ in correlator (1).

Some examples of diagrams, corresponding to different operators
contribution to this sum rules, are shown on fig1: Fig.1a
correspond to unit operator contribution (first term in (3)),

Fig.1b- to gluon condensate contribution (third term in (3)),
Fig.1c- to four-quark operator contribution (second term in (3)),
Fig.1d- to $d=8$ contribution (last term in (3)), Fig.1e- to quark
condensate contribution (first term in (4)), Fig.1f- to
quark-gluon  condensate  contribution, $\langle 0 \vert \bar{\psi}
(\lambda^n/2) G^n_{\mu \nu} \sigma^{\mu \nu} \psi\ \vert 0
\rangle=m_0^2\langle \bar{\psi} \psi\rangle$, (it contribution
found to be 0), Fig.1g- to $d=7$ contribution (last term in (4)),
Fig.1h- to $d=9$ condensate contribution (second term in (4)).

Factorization hypothesis for all operator with $d>5$ was used,
that's why,  a number of operators of dimension 6 (like $\langle 0
\vert \bar{\psi} \Gamma \psi \bar {\psi} \Gamma \psi \vert 0
\rangle$, where $\Gamma= I, \gamma^5,~ \gamma^{\mu},~ \gamma^{\mu}
\gamma^5, ~\sigma^{\mu \nu}$) and operator with higher dimension
were expressed in terms of $a$, $b$ or $m_0^2a$. One should note,
that contribution of $d=6$ operators in sum rule (3)  is rather
large (and larger than operator $d=4$) but the operators of d=8
are small, so the standard condition that higher terms in OPE in
sum rules should be less then $30\%$ is fulfilled at $d>6$ (see
$\cite{1,2}$).

Some years later perturbative correction to bare loop, quark
condensate and four-quark condensate was calculated (in
$\overline{MS}$ scheme) (see $\cite{3}$ and $\cite{4}$). It is
necessary to note, that in calculation of perturbative corrections
for $d=6$ four quark operator $\langle 0 \vert \bar{\psi} \Gamma
\psi \bar {\psi} \Gamma \psi \vert 0 \rangle$, one should not use
factorization hyphothesis from very beginning, but renormalize
each operator separetely, taking in acoount mixing and only in
final answer use factorization hyphothesis. This procedure was
done by authors of see $\cite{3}$ very carefully, and for
numerical estimations one can use the following relation, (based
on result see $\cite{3}$)

$$ M^6 E_2 (s_0/M^2)L^{-4/9} \Biggl [ 1+ \Biggl (\frac{53}{12}
+\gamma_E\Biggr ) \frac{\alpha_s(M^2)}{\pi} \Biggr ] +
\frac{M^2}{4}b E_0(s_0/M^2) +$$

\be + \frac{4}{3} a^2 \Biggl (1+\frac{\alpha_s(M^2)}{\pi}
(-1/9+\gamma_E/3)\Biggr ) -\frac{1}{3} \frac{a^2m^2_0}{M^2} =
\overline{\lambda}^2_N e^{-m^2/M^2}\ee

\be 2aM^4 E_1(s_0/M^2) (1+\frac{3}{2}\alpha_s/\pi) +\frac{272}{81}
\frac{\alpha_s a^3}{\pi M^2} -\frac{1}{12} ab
=m\overline{\lambda}_N^2 e^{-m^2/M^2} \ee

(The results of paper $\cite{5}$ coincide with $\cite{3}$ for bare
loop and quark condensate and for $d=6$ four quark condensate
differ only in non-logarithmic term in $\alpha_s$ correction. This
difference can be connected with slightly difference factorization
scheme).

In eq.(5,6) $\alpha_s(M^2)$ should be accounted up to second  term
of perturbative expansion, but the better is to use well-known
renormgroup relation for $ln~Q^2/M^2 =
-\int\limits^{\alpha_s(Q^2)/\pi}_{\alpha_s(\mu^2)/\pi}
\frac{d\alpha_s}{\pi\beta (\alpha_s/\pi)}$ where
$\beta(x)=\sum_0\beta_n x^{n+2}$, and terms $\beta_n$ up to fourth
one can find in $\cite{6,7}$ (for review see, for example,
$\cite{8}$) and fix normalization point $\mu$ at Z-bozon mass,
where $\alpha_s$ is well-known.

These results, on which eq. (5) is based, were obtained more them
15 years ago. From this time the situation with parameters changes
drastically (and that's why reanalysis of baryon sum rules
 seems to be necessary). First of all, $\Lambda_{QCD}$ became
 about two time larger (the modern value is about $300~MeV$).
Also large uncertainity appear for the value of quark condensate.
The
 well-known estimation of quark condensate, based on
 Gel-Mann-Oakes-Renner relation lead to value of quark condensate
 $\langle \bar{\psi}\psi \rangle= -(243MeV)^3$ (see for example
 review $\cite{8}$). The
 normalization point for it supposed to be about $1 GeV$, (to be
 able use this relation in QCD calculation),
 which is doubtfull for relation obtained as low-energy theorem.
 In this case renorminvariant value $\bar{a}^2=\alpha_s a^2$ is
 equal
\be \bar{a}^2=0.26 GeV^6 \ee

but accuracy is about $50\%$ (see discussion in $\cite{8}$).
 From other side, there are estimations of quark
condensate from sum rules  $\tau$ lepton decays (for review see
$\cite{8}$) which lead to much larger value

\be \bar{a}^2=0.47 \pm0.14~GeV^6 \ee.

In $\cite{8}$ it was supposed that real value should be close to
$\bar{a}^2=0.34 ~GeV^6$ which is close to the boundary of this two
equations (upper for (6) and lower for(7)).

Finally, the value of gluon condensate $ b=(2\pi)^2 \langle
(\alpha/\pi) G^n_{\mu \nu}G^n_{\mu \nu} \rangle$ which usually
supposed to be $0.47~GeV^4$ was also changed last years as a
result of changing $\Lambda_{QCD}$ from one side and more precise
analysis of the sum rules with higher terms of pertuirbative
corrections with other. The modern estimation is (see review
$\cite{8}$)

\be b=0.35 \pm 0.28 \ee

so one can see that uncertainity is very large, and the
possibility that gluon condensate is equal to zero isn't excluded.

In all this reasons we in our analysis of sum rule should vary

\be 0.23 GeV^6< \bar{a}^2 < 0.34GeV^6,~~ 0<b<0.48GeV^4 \ee

The parameter $m_0^2$ we choose in standard way equal to
$0.8GeV^2$ (we don't vary it because terms proportional to this
parameter are rather small). We choose continuum threshold
$s_0=2.3GeV^2$, according $\cite{2}$. Numerical analysis show,
that variation of $s_0$ in the reasonable region
$2.1GeV^6<S_0<GeV^6$ give less than $10\%$ variation in eq.(5,6)

Let us now discuss the sum rules (5,6). We see that we have two
different equations for the $\overline{\lambda_N}^2$. That's mean
that the ratio $R$ of this two equations should be close to unity
and it deviation from unity should indicate the accuracy of the
sum rule. But one should note that this sum rules has very high
accuracy itself due the fact that large number of the OPE series
are taken in account and also perturbatuve correction are
accounted (if we for a moment forget about accuracy of
condensates, i.e supposed that condensates are fixed). Estimations
show, that accuracy of sum rules (5,6) are about $10\%$. An
additional argument of such high accuracy is Borel mass behaviour,
which is practically constant (see as an example fig.2, which we
shall discuss little later). That's why, from our point of view
the deviation of the ratio $R$ from 1 more than $15\%$ indicates
that choose of parameters from (10) is incorrect. So we can
restrict the area of variation of $b$ and $\bar{a}^2$. On Fig.3a,b
the Borel mass dependence of ratio $R$ is shown for different
choices of $b$ and $\bar{a}^2$. Thin, thick and dash lines
correspond to $b$=0.48, 0.24 and 0 $GeV^4$, fig.3a correspond to
$\bar{a}^2=0.34~GeV^6$, and fig. 3b - to $\bar{a}^2=0.23~GeV^6$.
One can see, that value $b$=0 should be excluded, $b$=0.24 $GeV^4$
is allowed only at $\bar{a}^2=0.23~GeV^6$, and $b$=0.48 $GeV^4$ is
good in whole region of  $\bar{a}^2$, and for
$\bar{a}^2=0.23~GeV^6$ the agreement is the best.

Result for nucleon coupling $\bar{\lambda_N}^2$ for
$\bar{a}^2=0.23~GeV^6$ and two values of $b =0.24,~ 0.48~GeV^4$
are shown on Fig.2 (thick and thin line correspondingly). One can
see that Borel mass dependence is almost constant, and estimated
accuracy is less than $15\%$. Finally we come to following
conclusions,

1. $\bar{\lambda_N}^2=~2.3~GeV^6$ with accuracy about $15\%$,

2. the reasonable lower value of gluon condensate is $0.24
~CeV^4<b$

3. $\bar{a}^2$ can vary in region from $0.23~GeV^6$, which
correspond to standard value of quark condensate (see eq. (7) up
to $0.34~GeV^6$, which is close to lower limit, obtained from sum
rules for $\tau$ lepton decay.

\newpage

{\bf Acknowledgement}

\vspace{3mm}

The author is thankful to B.L.Ioffe for many useful discussions
and advises. The research described in this publication was made
possible in part, by INTAS Grant 2000, Project 587 and by the
Russian Found of Basic Research, Grant No. 00-02-17808.

\newpage

\begin{figure}
\epsfxsize=10cm
\epsfbox{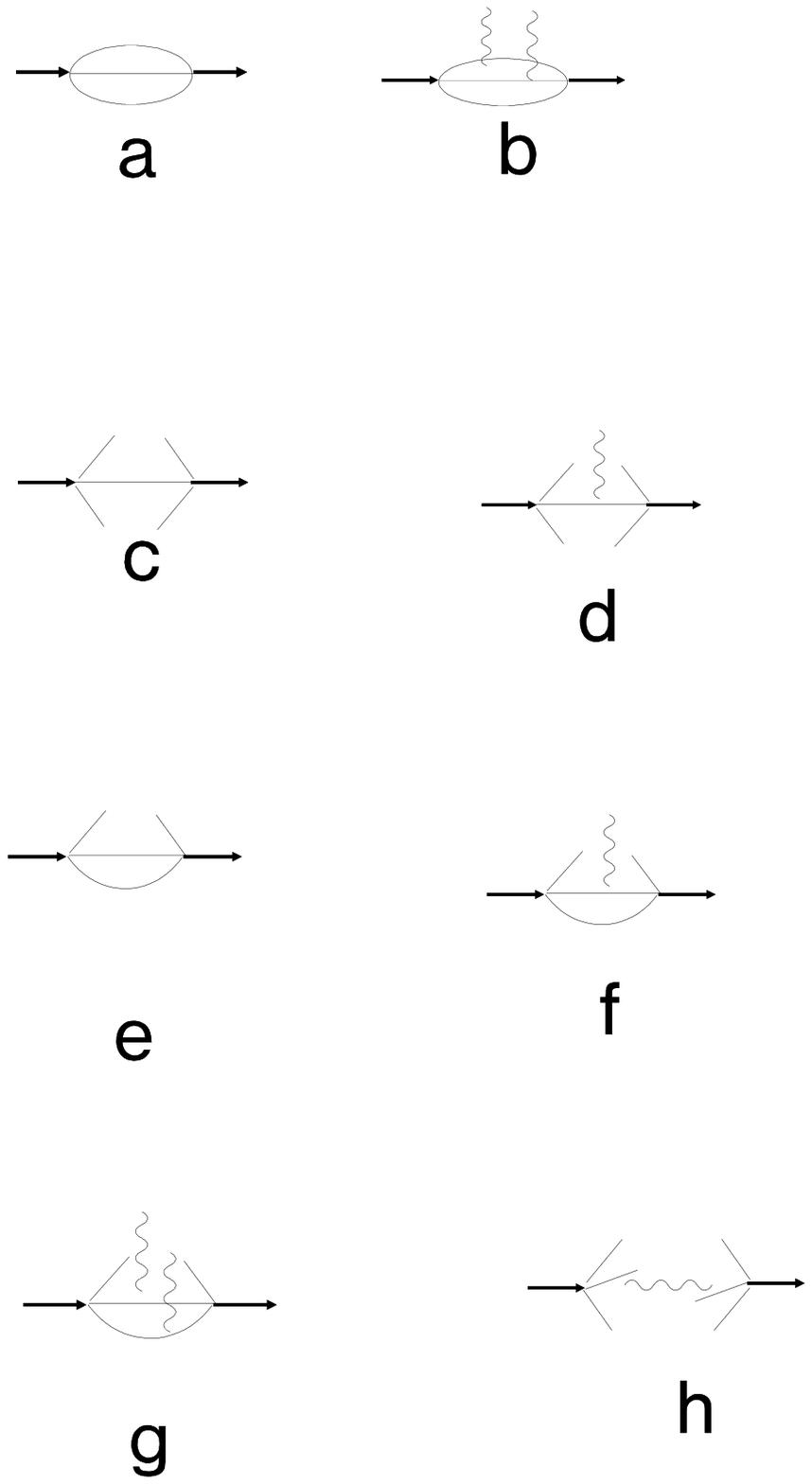}
\caption{Examples of diagram
for different operator contribution, wavy lines correspond to
gluon.}
\end{figure}

\begin{figure}
\epsfxsize=10cm \epsfbox{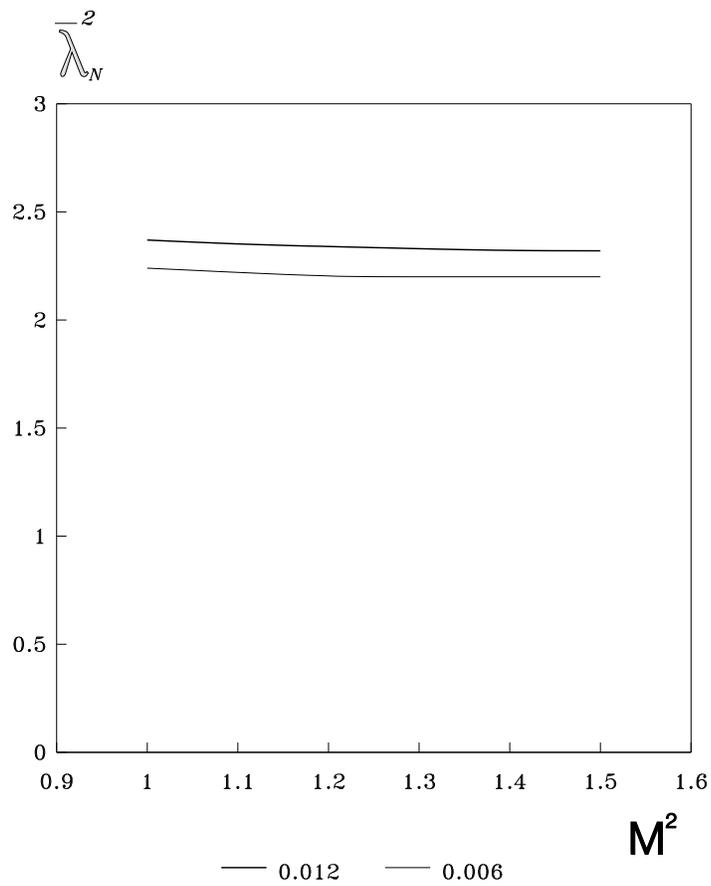} \caption{Dependence of proton
coupling constant  $\bar{\lambda_N}^2$  from Borel mass for gluon
condensate $b$=0.24 and 0.48$GeV^4$}
\end{figure}

\begin{figure}
\epsfxsize=10cm \epsfbox{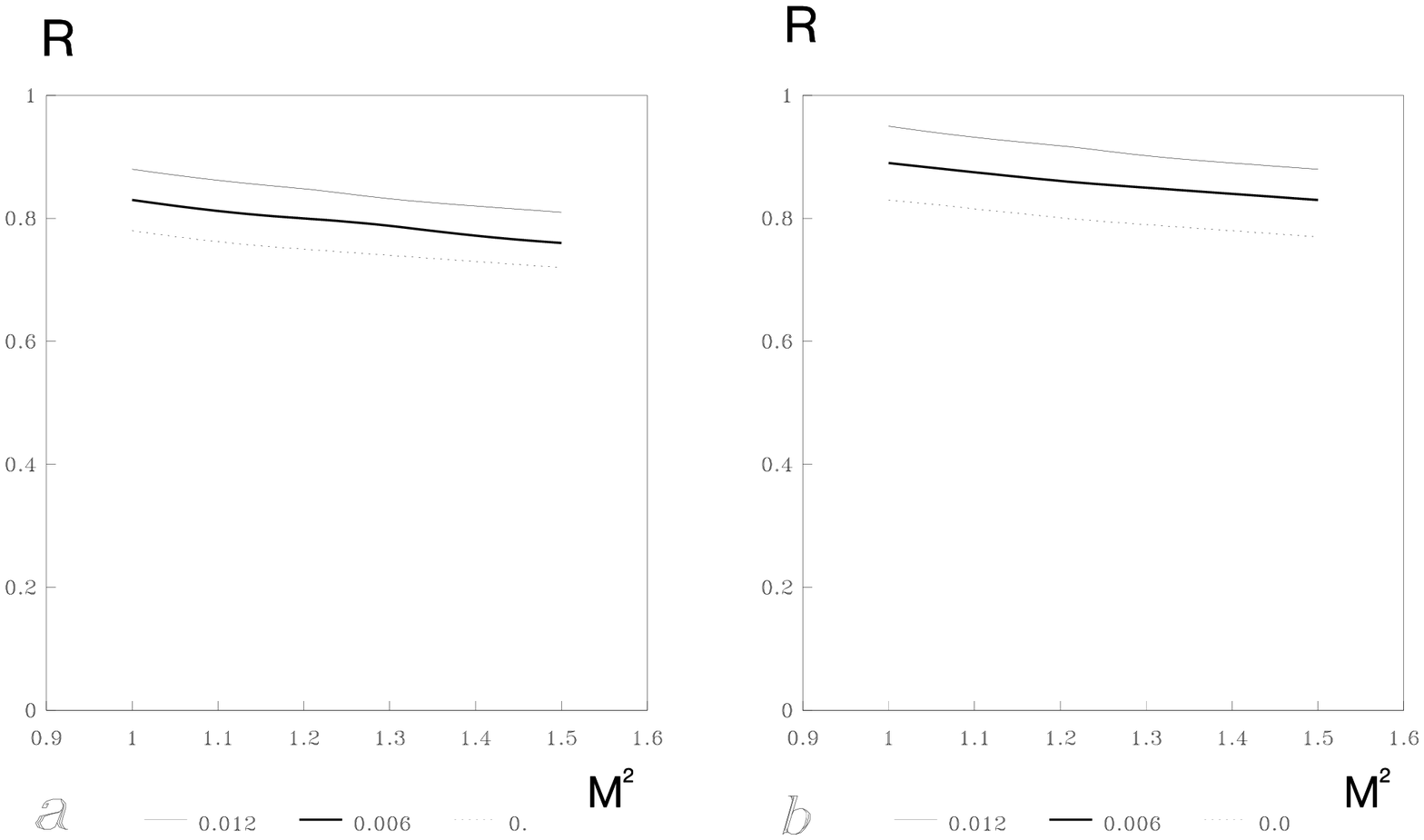} \caption{Dependence of ratio $R$
from borel mass for various set of gluon and quark condensates
from (10)}
\end{figure}


\newpage

\begin{thebibliography}{99}

\bibitem{1} V.M.Belyaev, B.L.Ioffe, Sov.Phys.JETP.  {\bf 56} (1982)
493.
\bibitem{2} B.L.Ioffe, A.V. Smilga, Nucl.Phys. {\bf B232} (1984)
109.
\bibitem{3} M. Jamin HD-THEP-88-19 (1988).
\bibitem{4} M. Jamin Z.Phys. {\bf C37} (1988) 635.
\bibitem{5} A.A.Ovchinnikov, A.A.Pivovarov, L.R. Surguladze, Jad.Phys,
 {\bf 40} (1988) 562.
\bibitem{6} S.A.Larin, J.A.M.Vermaseren Phys.Lett,{\bf B303} (1993) 334.
\bibitem{7} T.van Ritbergen, J.A.M.Vermaseren and S.A.Larin,
J.A.M.Vermaseren Phys.Lett,{\bf B400} (1997) 395
\bibitem{8} B.L.Ioffe, Jad.Phys,{\bf 66} (2003) 32.

\end{thebibliography}
\end{document}